\documentclass[apjpt4]{aastex}
\shorttitle{SN 2013ej}
\shortauthors{Fang Huang et al.}

\usepackage{graphicx}
\usepackage{longtable}

\begin{document}

\title{SN 2013ej in M74: A Luminous and Fast-declining \\ Type II-P Supernova}
\author{Fang Huang\altaffilmark{1,2}, Xiaofeng Wang\altaffilmark{2}, Jujia Zhang\altaffilmark{3,4,5}, Peter J. Brown\altaffilmark{6}, Luca Zampieri\altaffilmark{7}, Maria Letizia Pumo\altaffilmark{8}, Tianmeng Zhang\altaffilmark{9}, Juncheng Chen\altaffilmark{2}, Jun Mo\altaffilmark{2}, Xulin Zhao\altaffilmark{2}}
\altaffiltext{1}{Department of Astronomy, Beijing Normal University, Beijing 100875, China; \mbox{huangfang@mail.bnu.edu.cn}}
\altaffiltext{2}{Physics Department and Tsinghua Center for Astrophysics, Tsinghua University, Beijing 100084, China; wang\_xf@mail.tsinghua.edu.cn}
\altaffiltext{3}{Yunnan Observatories, Chinese Academy of Sciences, Kunming 650011, China}
\altaffiltext{4}{Key Laboratory for the Structure and Evolution of Celestial Objects, Chinese Academy of Sciences, Kunming 650011, China}
\altaffiltext{5}{University of Chinese Academy of Sciences, Chinese Academy of Sciences, Beijing 100049, China}
\altaffiltext{6}{George P. and Cynthia Woods Mitchell Institute for Fundamental Physics \& Astronomy, Texas A. \& M. University, Department of Physics and Astronomy, 4242 TAMU, College Station, TX 77843, USA}
\altaffiltext{7}{INAF-Osservatorio Astronomico di Padova, Vicolo dell'Osservatorio 5, 35122 Padova, Italy}
\altaffiltext{8}{INAF-Osservatorio Astronomico di Palermo `Giuseppe S. Vaiana', Piazza del Parlamento 1, 90134 Palermo, Italy}
\altaffiltext{9}{Key Laboratory of Optical Astronomy, National Astronomical Observatories, Chinese Academy of Sciences, Beijing 100012, China}

\begin{abstract}
We present extensive ultraviolet, optical, and near-infrared observations of the type IIP supernova (SN IIP) 2013ej in the nearby spiral galaxy M74. The multicolor light curves, spanning from $\sim$ 8--185 days after explosion, show that it has a higher peak luminosity (i.e., M$_{V}$ $\sim$$-$17.83 mag at maximum light), a faster post-peak decline, and a shorter plateau phase (i.e., $\sim$ 50 days) compared to the normal type IIP SN 1999em. The mass of $^{56}$Ni is estimated as 0.02$\pm$0.01 M$_{\odot}$ from the radioactive tail of the bolometric light curve. The spectral evolution of SN 2013ej is similar to that of SN 2004et and SN 2007od, but shows a larger expansion velocity (i.e., $v_{Fe II} \sim$ 4600 km s$^{-1}$ at t $\sim$ 50 days) and broader line profiles. In the nebular phase, the emission of H$\alpha$ line displays a double-peak structure, perhaps due to the asymmetric distribution of $^{56}$Ni produced in the explosion. With the constraints from the main observables such as bolometric light curve, expansion velocity and photospheric temperature of SN 2013ej, we performed hydrodynamical simulations of the explosion parameters, yielding the total explosion energy as $\sim$0.7$\times$ 10$^{51}$ erg, the radius of the progenitor as $\sim$600 R$_{\odot}$, and the ejected mass as $\sim$10.6 M$_{\odot}$. These results suggest that SN 2013ej likely arose from a red supergiant with a mass of 12--13 M$_{\odot}$ immediately before the explosion.
\end{abstract}


\section{Introduction}
\label{sect:intro}
Supernovae are violent explosions of stars at the end stage of their lives. Based on the explosion mechanisms, supernovae are classified into two main classes: thermonuclear explosions of white dwarf stars and core-collapse explosions of massive stars. The latter can be further subdivided as type II, Ib, and Ic SNe in terms of diverse features of hydrogen and/or helium lines in their optical spectra (\citealt{Filippenko1997, Turatto2007}). SNe II are defined by the presence of prominent hydrogen lines in the spectra; and type IIP SNe (SNe IIP) are specifically named after a long period of roughly constant luminosity (a plateau). SNe IIP are the most common subtype of stellar explosions, comprising nearly 60\% of all core-collapse (CC) SNe \citep{Li2011}.

SNe IIP are believed to emerge from CC explosions of relatively massive stars (i.e., M$_{ZAMS}$ $\ge$ 8 M$_{\odot}$) which retain most of the hydrogen-rich envelope before explosion. Theoretical models of stellar evolution suggest that the initial mass of the progenitors lies between 8 and 25 M$_{\odot}$ \citep{Heger2003}, while hydrodynamical modeling of a number of events gives a larger mass in the range of 15--25 M$_{\odot}$ \citep{Utrobin2009}. On the other hand, observational constraints on the properties of SNe IIP progenitors can be obtained by examining pre-explosion images \citep{Li2007, Smartt2009} or modeling of late-time spectra \citep{Jerkstrand2012}. These analyses unanimously suggest that SNe IIP arise from red supergiant (RSG) stars with initial masses of 8.5--16.5 M$_{\odot}$. The facts that no RSG stars with initial masses of 17--30 M$_{\odot}$ have been identified as the progenitors of SNe IIP in the pre-explosion images, also called the `RSG problem', may be explained by an underestimation of the progenitor luminosity due to extra circumstellar extinction \citep{Walmswell2012} or that the massive RSG stars tend to produce explosions of other core-collapse SNe such as SNe IIL \citep{Faran2014a}.

SNe IIP have been thought to have similar photometric behaviors, with a rapid decline in the first 30 days after maximum due to cooling down of the external layers of the progenitor, followed by a nearly constant, or so-called `plateau' phase which lasts for about 50--80 days \citep{Anderson2014}. The plateau phase is sustained by the recombination of ionized hydrogen. After that, the light curve experiences a transition to a linear decline (in magnitude) in the nebular phase when the supernova is powered by the radioactive decay. There is increasing evidence that SNe IIP show significant scatter in peak luminosity, duration of plateau phase, and expansion velocity of ejecta \citep{Hamuy2003, Faran2014b, Spiro2014, Zhang2014a}. SN 2013ej, exploded in M74 at a distance of about 9.7 Mpc (see \S 3.1), provides us a good chance to examine in detail the explosion physics and progenitor properties of SNe IIP.

\cite{Valenti2014} presented the first-month photometric and spectroscopic observations of SN 2013ej, confirming its membership of the class of SNe IIP. They show that this type II-P SN has an unusually long rise time ($\sim$ 10 days) before entering into a plateau phase. The $BVRI$ photometry has been presented by \cite{Richmond2014}, showing photometric features similar to some SNe IIP. However, the early-time spectropolarimetric observations taken at a few days after explosion reveal that this supernova has a strong polarization that is unusual for a normal SN IIP \citep{Leonard2013}, implying substantial asymmetries in the outer ejecta of SN 2013ej at early times.

In this work, we present extensive optical, ultraviolet (UV), and near-infrared (NIR) observations of SN 2013ej, with an attempt to put better constraints on the properties of this type II-P supernova. This paper is organized as follows: in Section \ref{sect:obs} we describe the observation and data reduction. Distance and extinction are discussed in Section \ref{sect:disex}. In Section \ref{sect:photometry} and \ref{sect:spectra}, we present the photometric and spectroscopic evolutions of SN 2013ej. Discussions are given in Section \ref{sect:discuss} and we summarize in Section \ref{sect:sum}.

\section{Observation and data reduction}  \label{sect:obs}
SN 2013ej was discovered by Lick Observatory Supernova Search (LOSS) on 2013 July 25.45 (UT is used throughout this paper) at an unfiltered magnitude of 13.5 mag \citep{Kim13} in the nearby spiral galaxy Messier 74 (M74 or NGC 0628), where the type Ic supernova SN 2002ap \citep{Mazzali2002} and the type II-P supernova SN 2003gd \citep{Hendry2005} were also recorded. SN 2013ej was located at $\alpha$ = 01h36m48s.16 and $\delta = 15\degr45 \arcmin 31 \farcs0$, 92.5\arcsec east and 135\arcsec south of the center of M 74 (see Figure \ref{fig:M74}). An optical spectrum obtained on July 27.73 suggests that it is a core-collapse supernova at a relatively young phase, showing a moderately blue contnuum with weak Balmer emission lines \citep{Valenti2013}. Extensive follow-up observations have been carried out for this object immediately after the discovery.

\subsection{Photometry}  \label{subsect:obs.photo}
\subsubsection{Ground-based Optical Observations}
Ground-based optical ($UBVRI$) photometry of SN 2013ej was obtained with the Lijiang 2.4-m telescope (hereafter LJT) of Yunnan Observatories \citep{Zhang2014b} and the 0.8-m Tsinghua University-NAOC telescope (hereafter TNT) at Xinglong Observatory in China \citep{Wang2008, Huang2012}. The LJT is equipped with the Yunnan Faint Object Spectrograph and Camera (YFOSC) observation system which works in both imaging and long-slit spectroscopic modes.

\begin{figure}
\centering
\includegraphics[height=14cm,trim=250 30 20 20]{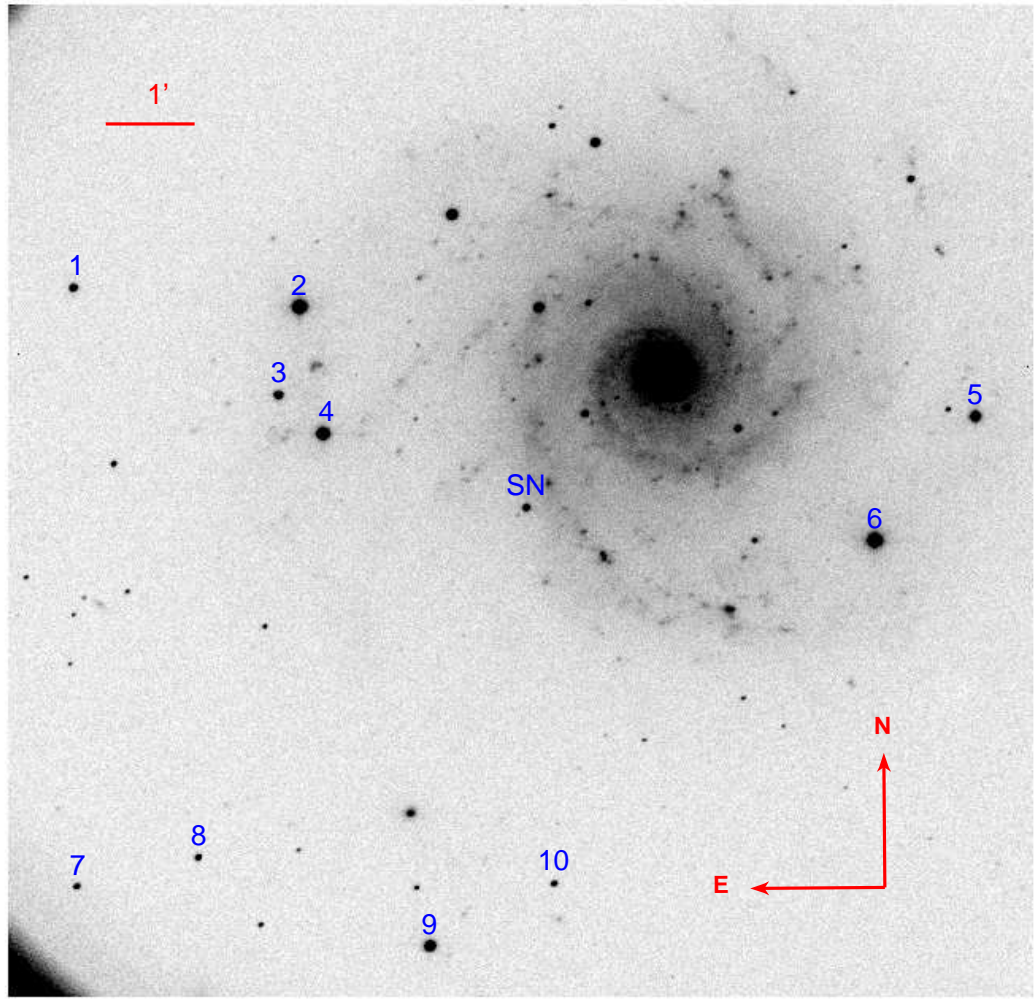}
\caption{SN 2013ej in M74. The $R$-band image was taken on November 7, 2013 by the 80-cm Tsinghua-NAOC telescope. The location of the SN and 10 local reference stars are marked. North is up, and east is to the left.}
\label{fig:M74}
\end{figure}

Pre-processing procedures, including bias subtraction, flat fielding, cosmic ray removal, were carried out by standard IRAF\footnote{IRAF is short for Image Reduction and Analysis Facility supported by the National Optical Astronomy Observatories (NOAO). NOAO is operated by the Association of Universities for Research in Astronomy (AURA), Inc. under cooperative agreement with the National Science Foundation.} routines. As SN 2013ej was very bright and exploded in a faint region of the host galaxy (see Figure \ref{fig:M74}), we did not apply the technique of template subtraction in the photometry. The aperture photometry method was used to obtain the instrumental magnitudes of SN 2013ej and 10 local standard stars as shown in Figure \ref{fig:M74}. Instrumental magnitudes were then converted into the standard Johnson $UBV$ \citep{Johnson1966} and Kron-Cousins $RI$ \citep{Cousins1981} systems, with the color terms and zeropoints determined via  a series of Landolt \citep{Landolt1992} standards observed on the photometric nights. The magnitudes of the local comparison stars are shown in Table \ref{tab:localstar}, and the final $UBVRI$ magnitudes of SN 2013ej with flux calibrations are presented in Table \ref{tab:photsn}.

\subsubsection{Optical/UV Observations from SWIFT UVOT}
The Swift UVOT (Roming et al. 2005) observations of SN 2013ej started from 2013 July 30.97, spanning for approximately 130 days (PIs: Chakraborti, Sokolovsky, \& Brown). The photometric observations of SN 2013ej were taken in three UV filters ($uvw$1, $uvm$2 and $uvw$2) and three optical filters ($u$, $b$, and $v$). The first-epoch observation showed that most of the UVOT filters are near or above their saturation limit due to the brightness of this supernova \citep{Margutti2013}. To avoid this effect on the measurements, some of the subsequent observations were done in a mode to decrease the frame readout time and reduce the effect of coincidence loss (see \citealt{Poole2008} for details). The early-time UVOT photometry, covering about 2 weeks from the explosion, was reported by \citet{Valenti2014}. Here we present the full set of $Swift$ UVOT photometric results of SN 2013ej in Table \ref{tab:uvot}, which are available from the Swift Optical/Ultraviolet Supernova Archive (SOUSA; Brown et al. 2014). These magnitudes are measured after applying subtraction of the galaxy template obtained with the Swift/UVOT during the period 2007-2008. Note that the template subtraction is performed using the UVOT flux (or count rates) rather than the images themselves, as described in Brown et al.(2014).

\subsubsection{NIR Observations from NTT SOFI}
The near-infrared observations of SN 2013ej were obtained with the 3.58-m ESO New Technology Telescope (NTT) equipped with the SOFI (Son of ISAAC) camera as part of the Public European Southern Observatory Spectroscopic Survey of Transient Objects (PESSTO, \citealt{Smartt2014}). Images were acquired through $JHK_{s}$ filters at 7 epochs spanning from $\sim$+24~d to +185~d after explosion. The NIR images were processed, including bias subtraction, flat-field correction, sky subtraction, alignment, combination of dithered frames and aperture photometry. We used the 2MASS point-source catalog \citep{Skrutskie2006} to calibrate the supernova. The calibrated $JHK$ magnitudes of SN 2013ej are reported in Table \ref{tab:nir}.

\subsection{Spectroscopy} \label{subsect:obs.spec}
A total of 13 low-resolution spectra of SN 2013ej were obtained with the YFOSC system on the LJT  and the BFOSC system on the 2.16-m telescope at NAOC Xinglong Observatory, spanning from +8 to +157 days since the explosion. A journal of spectroscopic observations of SN 2013ej is given in Table \ref{tab:speclog}.

All spectra were reduced in the standard IRAF environment. The routine preprocessing procedures include the bias subtraction, corrections for the flat field, and removal of the cosmic rays. One-dimensional spectra were extracted using $apall$ task. The He/Ne lamp and Fe/Ar lamp were used for the wavelength calibration of the YFOSC and BFOSC spectra, respectively. To get a better flux calibration, the spectrophotometric standard star Feige 15 at a similar air mass was observed on the same night. In addition, the flux-calibrated spectra were corrected for the continuum atmospheric extinctions obtained at Xinglong and Li-Jiang Observatories, with further removals of the telluric lines.

\section{Distance and Extinction} \label{sect:disex}
\subsection{Distance to SN 2013ej} \label{subsect:distance}
By far, there are existing estimates of the distances to M74 by many methods except for the Cepheids. Based on the HST images and Tip of the Red Giant Branch method, \cite{Jang2014} derived a distance to M74 as $30.04\pm0.04{\rm (random)}\pm0.12{\rm (systematic)}$ mag, corresponding to a linear distance of $10.19\pm0.14 {\rm (random)}\pm0.56{\rm (systematic)}$ Mpc. The distance estimated by the Tully-Fisher relation is $9.7\pm1.8$ Mpc, or $\mu=29.93\pm0.40$ mag \citep{Tully1988}. Using the Standardized Candle Method (\citealt{Hamuy2002, Hamuy2003}), \cite{Hendry2005} estimated a distance of $9.6\pm2.8$ Mpc with the $V$- and $I$-band observations of SN 2003gd, while \cite{Olivares2010} gave an updated value of 9.9$\pm1.3$ Mpc with the same method. \cite{Herrmann2008} obtained a distance of 8.6$\pm$0.3 Mpc using the planetary nebula luminosity function. These estimates agree with each other within the quoted errors. However, a smaller value, i.e., $7.8\pm0.4$ Mpc ($\mu=29.46\pm0.11$ mag), was reported using the brightest blue stars \citep{Sharina1996}. This value was not adopted in our analysis according to the rejection of 3-$\sigma$ outlier. The rest estimates give an averaged value of 9.6$\pm$0.7 Mpc ($\mu=29.91\pm0.16$ mag), which is adopted in the following analysis.

\subsection{Extinction} \label{subsection:extinction}
The Galactic reddening towards the line-of-sight direction of SN 2013ej is $E(B-V)_{Gal}$=0.06 mag \citep{SF11}, corresponding to an extinction of 0.19 mag in $V$ for $R_V$=3.1\citep{Cardelli1989}. The host-galaxy component can be estimated from spectroscopic and photometric methods. We examined our 13 low-resolution spectra but failed to detect any significant signature of Na I D absorptions due to the host galaxy, which indicates that SN 2013ej may suffer negligible reddening in the host galaxy. This is further evidenced by the non-detection of such Na I lines in the high-resolution spectrum of SN 2013ej \citep{Valenti2014}. However, it is worthwhile to point out that there is a large scatter between the strength of Na I absorption and the amount of line-of-sight reddening (e.g., \citealt{Poznanski2012}).

As an alternative, we applied the photometric method to estimate the host galaxy reddening for SN 2013ej. The $V-I$ color has been proposed as a better reddening indicator for SNe IIP because it exhibits more uniform properties (especially during the plateau phase) and it is also expected to be less sensitive to the metallicity effects. For example, \cite{Hamuy2004} proposed a method to estimate the reddening of SNe IIP by comparing the $V-I$ color curve with that of the well-studied template such as SN 1999em. A host-galaxy reddening of $E(B-V)_{host}=0.06\pm0.06$ can be inferred for SN 2013ej assuming a reddening of $E(B-V)_{tot}=0.10\pm0.05$ mag for SN 1999em \citep{Baron2000}. \cite{Olivares2010} suggest that the host-galaxy reddening of SNe IIP can be derived from the $V-I$ color toward the end of the plateau phase (i.e., $\sim$ 30 days before the end of the plateau phase). For SN 2013ej, this phase corresponds to t $\sim$ 60 days after the explosion when the $V-I$ color is measured to be 0.74$\pm$0.07 mag. Based on the empirical relation, $A_V(V-I) = 2.518[(V-I)-0.656]$, this results in a host galaxy reddening of $E(B-V)_{host} = 0.07\pm$0.08 mag for $R_V$ = 3.1 after correcting for the Galactic reddening. Therefore, a total reddening of $E(B-V)_{tot}=0.12\pm0.06$ can be derived for SN 2013ej, which corresponds to a total line-of-sight extinction of $A_V= 0.37\pm$0.19 mag for R$_{V}$ = 3.1.

\section{Photometric Evolution} \label{sect:photometry}
\begin{figure}
\centering
\includegraphics{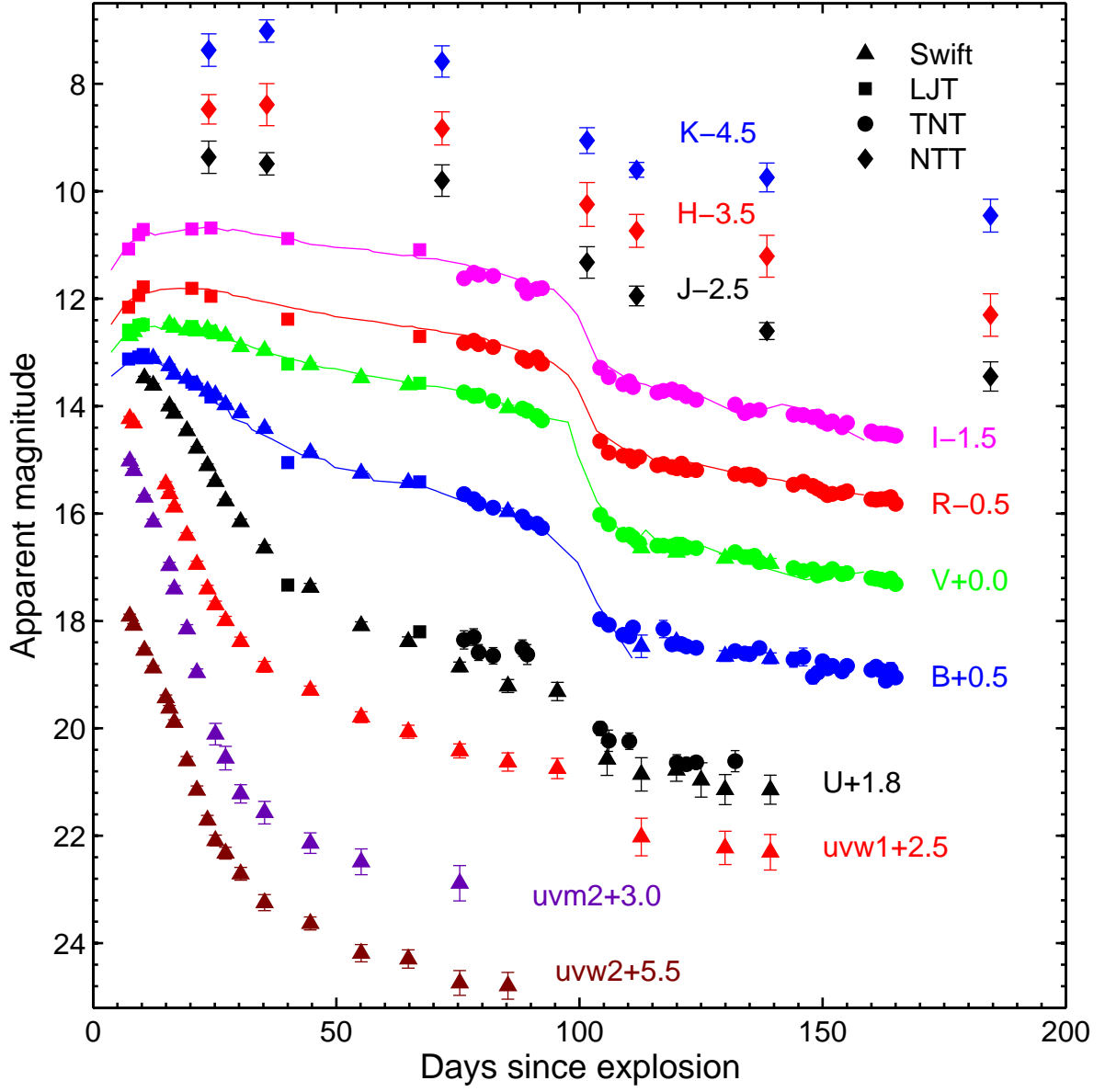}
\caption{The optical, ultraviolet and near-infrared light curves of SN 2013ej. The optical data from \cite{Richmond2014} are overplotted as solid lines for comparison. The light curves are shifted in magnitudes for better display. In $BVRI$ bands, the error bars are smaller than the symbols.}
\label{fig:lc}
\end{figure}

The UV-optical-NIR (UVOIR) light curves of SN 2013ej are shown in Figure \ref{fig:lc}. The data published by \cite{Richmond2014} are also overplotted for comparison. In the optical bands, the observations of the LJT and TNT agree well with those from $Swift$ and \cite{Richmond2014}. The UV light curves show much faster post-maximum evolution than those in the optical and NIR bands. Detailed analysis of the light curves and color evolutions is presented in the following subsections.

\subsection{UV Light Curves}
SN 2013ej is initially bright in the UV, then it undergoes a rapid linear decline over the first few days. The decline rate in $uvm$2 band is slightly steeper (0.27 mag day$^{-1}$) than that in $uvw$1 (0.18 mag day$^{-1}$) and $uvw$2 (0.24 mag day$^{-1}$), despite that the decay rate generally steepens at shorter wavelengths. \cite{Brown2007} explained this with the appearance of strong Fe II and Fe III lines in the $uvm$2 band. After t $\sim$ 30 days, the UV light curves flatten into a slowly declining plateau, which is similarly seen in normal SNe IIP \citep{Brown2009, Pritchard2014}. Two effects might contribute to this leveling off in the UV bands during the early nebular phase. One is related to the evolution of the photospheric temperature which slows down by one month after the maximum light. The other is likely due to that the UV emission at this phase is contaminated by the photons leaked from the optical tails of $uvw2$ and $uvw1$ filters.

\subsection{NIR Light Curves}
The NIR-band light curves of SN 2013ej have the same features as the VRI-band light curves, showing a plateau phase of nearly constant brightness that lasted for about 50 days (see \S4.3). In the J band, SN 2013ej gradually decreased its luminosity during the plateau phase, while in the H and K bands its luminosity initially increased and then decreased at similar times. This is somewhat different from the features seen in other SNe IIP such as SN 1999em \citep {Hamuy2001} and SN 2005cs \citep {Pastorello2009} where the luminosity increases monotonically close to (or until) the end of the plateau. For SN 2013ej, the degree of brightening increased with wavelength, and reached $>$ 0.35 mag in the K band. After entering the nebular phase, the NIR luminosity declines at a rate similar to that in the optical bands.

\subsection{Optical Light Curves}
\subsubsection{Apparent Light Curves }
SN 2013ej was first detected in a KAIT image taken on 2013 July 25.45 \citep{Kim13}, but the pre-discovery detection on 2013 July 24.13 was reported by C. Feliciano on the \textit{Bright Supernova}\footnote{http://www.rochesterastronomy.org/snimages/} website. This supernova was not detected on 2013 July 23.54 according to the images from the All Sky Automated Survey for Supernovae \citep{Shappee2013}, with a limiting magnitude of $V$ \textgreater 16.7 mag. Assuming that SN 2013ej exploded near July 23.5 (which is also adopted in \citealt{Valenti2014}), we find a rise time to the maximum light of 15.0$\pm$1.0 days in the $V$ band.

With the LJT, TNT, and $Swift$ light curves, we estimated the date of maximum light and peak magnitude in $BVRI$ bands by fitting a cubic polynomial to the points around maximum light. The results are listed in Table \ref{tab:maxpara}. It is found that SN 2013ej reached a $B$-band maximum brightness of 12.56 $\pm$0.04 mag on JD 2456507.23 (August 2.73, 2013), and it reached at the $V$-band maximum of 12.45$\pm$0.05 mag at about 5 days later. The obtained peak magnitudes are consistent with the results from \cite{Richmond2014} within the quoted error bars.

For normal SNe IIP, the optical light curves go through a plateau phase shortly after the maximum. In SN 2013ej, the plateau feature can be seen in the VRI bands but it is inapparent in the $B$ band; while the light curve in the U band shows a monotonic decline. This behavior is similarly seen in SN 2007od \citep{Inserra2011}. The plateau of SN 2013ej is relatively short compared to that of normal SNe IIP. For example, the $V$-band magnitude declines by $\sim$1.0 mag in the first 50 days from the peak, which is much larger than most  SNe IIP (see Figure 6 in \citealt{Faran2014a}). We note that the light curves at shorter wavebands tend to have larger decline rates, which is likely related to the rapid cooling of the photosphere as a result of line blanketing in bluer bands.

After the plateau phase, there is a steep decline in the brightness in all bands, indicating the transition from the photospheric phase to the nebular phase. In SN 2013ej, this phase begins at $\sim$ +90 days since the explosion and ends $\sim$ 20 days later. After t $\sim$ +110 days, the light curves decline linearly in all bands, with a decay rate of 1.30 ${\rm mag}\,(100{\rm d})^{-1}$ in the $V$ band. At this phase, the photometric evolution of SNe IIP is powered by radioactive decay of $^{56}$Co to $^{56}$Fe, and the expected decay rate is 0.98 ${\rm mag}\,(100{\rm d})^{-1}$, especially in the $V$ band. The larger decay rates seen in SN 2013ej indicate that the gamma rays are not efficiently trapped in the ejecta, likely suggestive of a smaller optical opacity and hence a less massive envelope for its progenitor star (see discussions in \S 6).

\subsubsection{Absolute Light Curves}
\begin{figure}
 \centering
\includegraphics{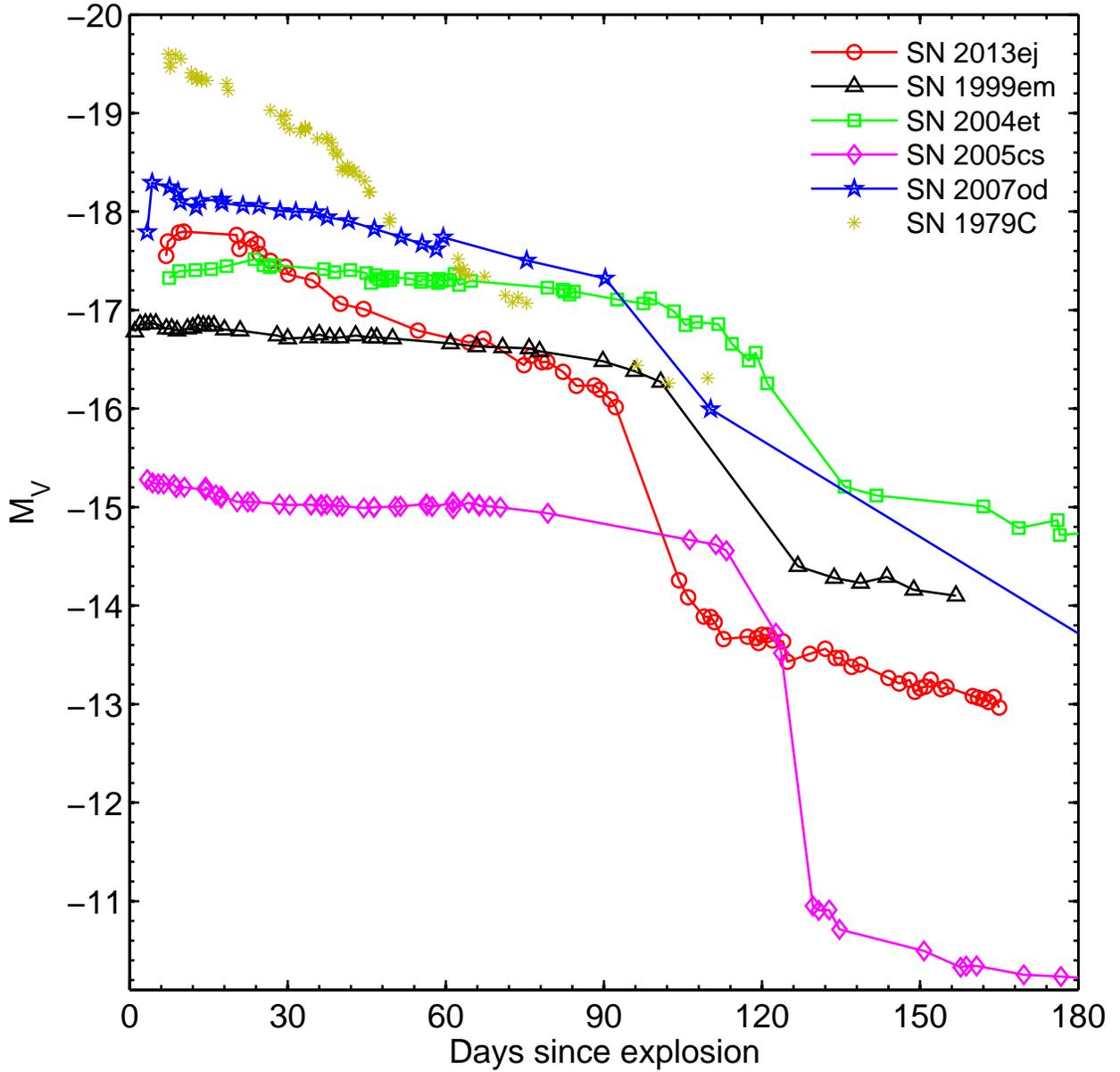}
\caption{Comparison of the $V$-band absolute light curve of SN 2013ej with those of other well-observed SNe IIP (SN 1999em, SN 2004et, SN 2005cs, and SN 2007od) and type IIL SN 1979C.}
\label{fig:absMv}
\end{figure}

In Figure \ref{fig:absMv}, we present the absolute $V$-band light curve of SN 2013ej, together with those of type II-P SN 1999em \citep{Leonard2002}, SN 2004et \citep{Sahu2006}, SN 2005cs \citep{Pastorello2009}, SN 2007od \citep{Inserra2011}, and the prototype of type II-L supernova 1979C \citep{deVau1981, Barbon1982}. All of these light curves have been corrected for extinction due to the Milky way and the host galaxies of the SNe. The comparison shows that the photometric evolution of SN 2013ej is very similar to that of SN 2007od. At around the peak, SN 2013ej has an absolute magnitude of M$_{V}^{max}$=$-$17.83$\pm$0.25 mag, which is fainter than SN 2007od and SN 1979C but much brighter than other SNe IIP of our sample, lying at the luminous end of SNe IIP according to the statistic results from \cite{Faran2014b}. After the maximum light, the luminosity of SN 2013ej declines at a rate similar to that of SN 1979C. During the plateau phase, the luminosity of SN 2013ej (i.e., M$_V(50)$= $-$16.96$\pm$0.33 mag) is close to that of SN 1999em. Note that the plateau feature is actually not well developed in SN 2013ej compared to that seen in some typical SNe IIP like SN 2004et and SN 1999em. To quantitatively describe the plateau feature of SN 2013ej, we fit its $V$-band light curve following the method defined in \cite{Anderson2014} and \cite{Olivares2010}. This yields an estimate of the plateau duration of $\sim$50 days. This duration is distinctly shorter than the typical value (i.e., $\sim$ 80 days) yielded for normal SNe IIP. It has been recently proposed that more luminous SNe IIP tend to have shorter plateau phases \citep{Poznanski2013}. A shorter plateau phase may indicate that the progenitor has a thinner hydrogen envelope before explosion so that the explosion energy cannot be sufficiently trapped.

\subsubsection{Color Evolution}  \label{subsect:colour}
\begin{figure}
 \centering
\includegraphics{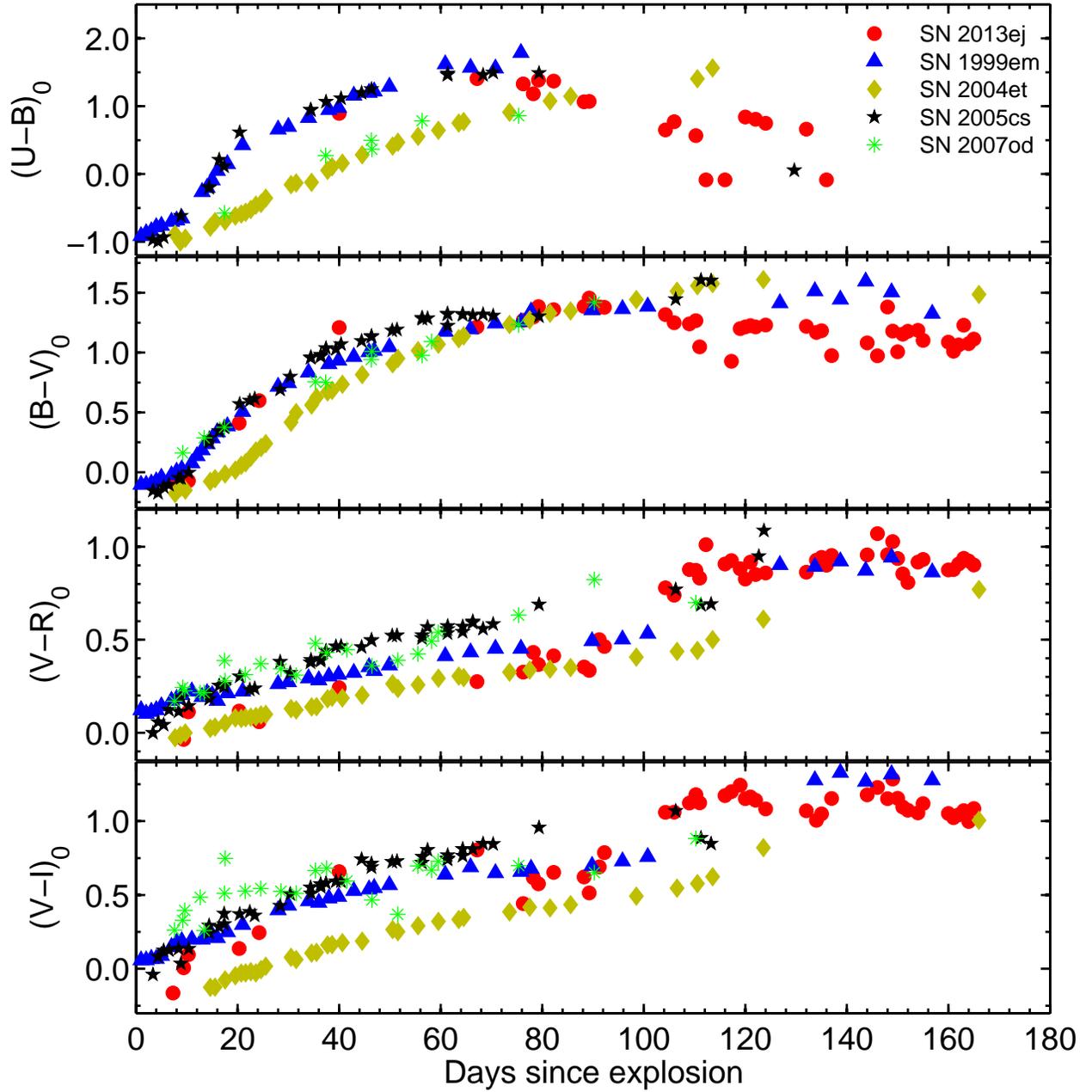}
\caption{Comparison of $UBVRI$ color curves between SN 2013ej and some representative SNe IIP.}
\label{fig:color}
\end{figure}

The $U-B$, $B-V$, $V-R$ and $V-I$ color curves of SN 2013ej are shown in Figure \ref{fig:color}, together with those of SNe 1999em, 2004et, 2005cs and 2007od. All of the color curves were corrected for the reddening. At early phase, SN 2013ej is quite blue due to high temperature from shock breakout, and it evolves redward rapidly especially in the $(U-B)_{0}$ and $(B-V)_0$ color as a result of the expansion and cooling of the photosphere. During the recombination phase, the colors remain almost constant. After t $\sim$ 120 days, the color curves of SN 2013ej show little evolution, similar to SN 1999em. One can see that $(U-B)_0$ and $(B-V)_0$ colors become redder earlier and at a pace faster than the $(V-R)_0$ and $(V-I)_0$ colors, which is likely due to the appearance of strong metallic lines at shorter wavelengths. The color evolution of SN 2013ej is overall very similar to that of the comparison SNe IIP.

\section{Optical Spectroscopy} \label{sect:spectra}
\begin{figure}
 \centering
 \includegraphics[]{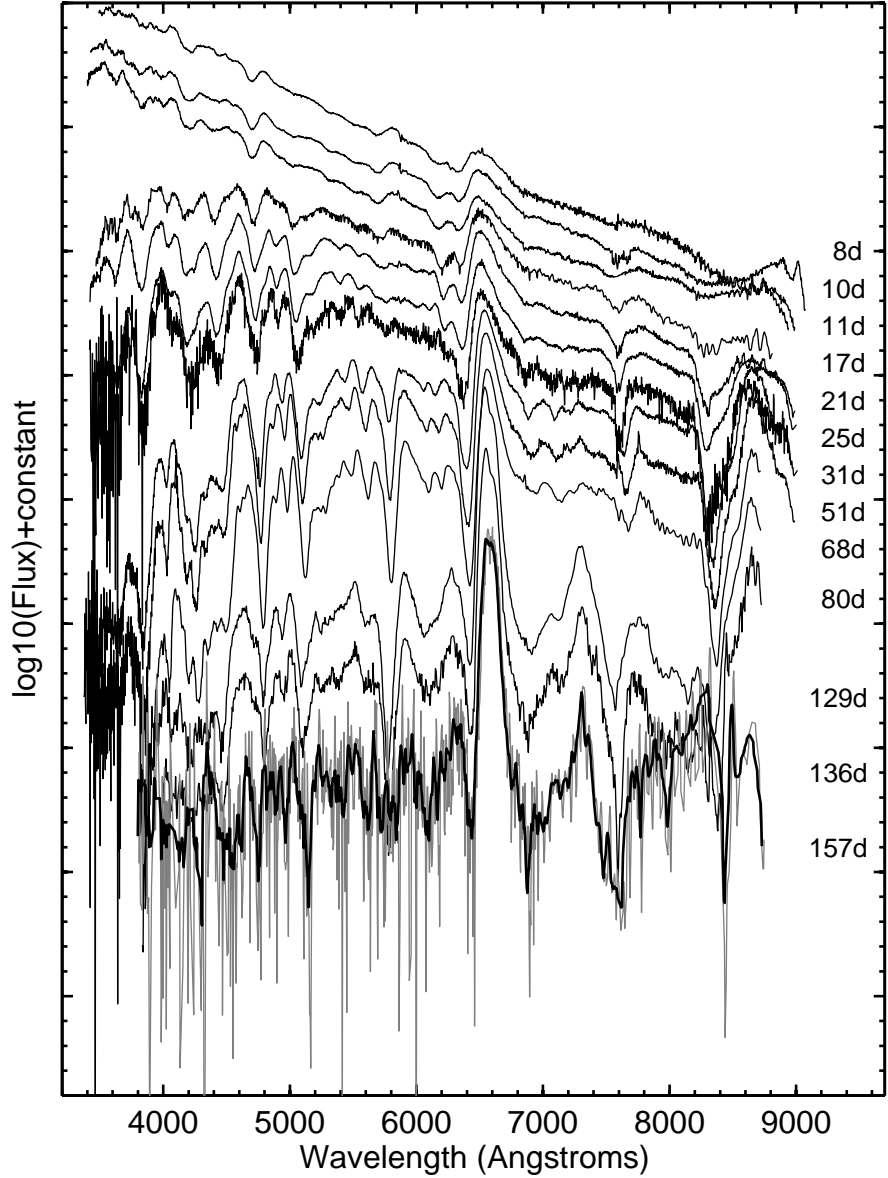}
 \caption{Optical spectral evolution of SN 2013ej. The spectra have been corrected for the redshift of the host-galaxy ($v_{hel} = 657\pm1$ km s$^{-1}$; \citealt{Lu1993}) and shifted vertically for better display. The last spectrum has been smoothed with a bin of about 25 \AA~ due to its low signal-to-noise ratio. The numbers on the right side mark the epochs of the spectra in days after explosion.}
\label{fig:spectra}
 \end{figure}

A total of thirteen optical spectra of SN 2013ej were obtained with the Lijiang 2.4-m telescope and the Xinglong 2.16-m telescope, spanning from t = +7.8 days to t = +157.0 days with respect to the explosion date. Figure \ref{fig:spectra} shows the complete spectral evolution, with the main spectral features being the development of line profiles of hydrogen, O I, and Ca II IR triplet when entering the nebular phase. The spectral evolution will be examined in detail in the following subsections.

\subsection{Evolution of the Spectra}
\begin{figure}
 \centering
 \includegraphics[]{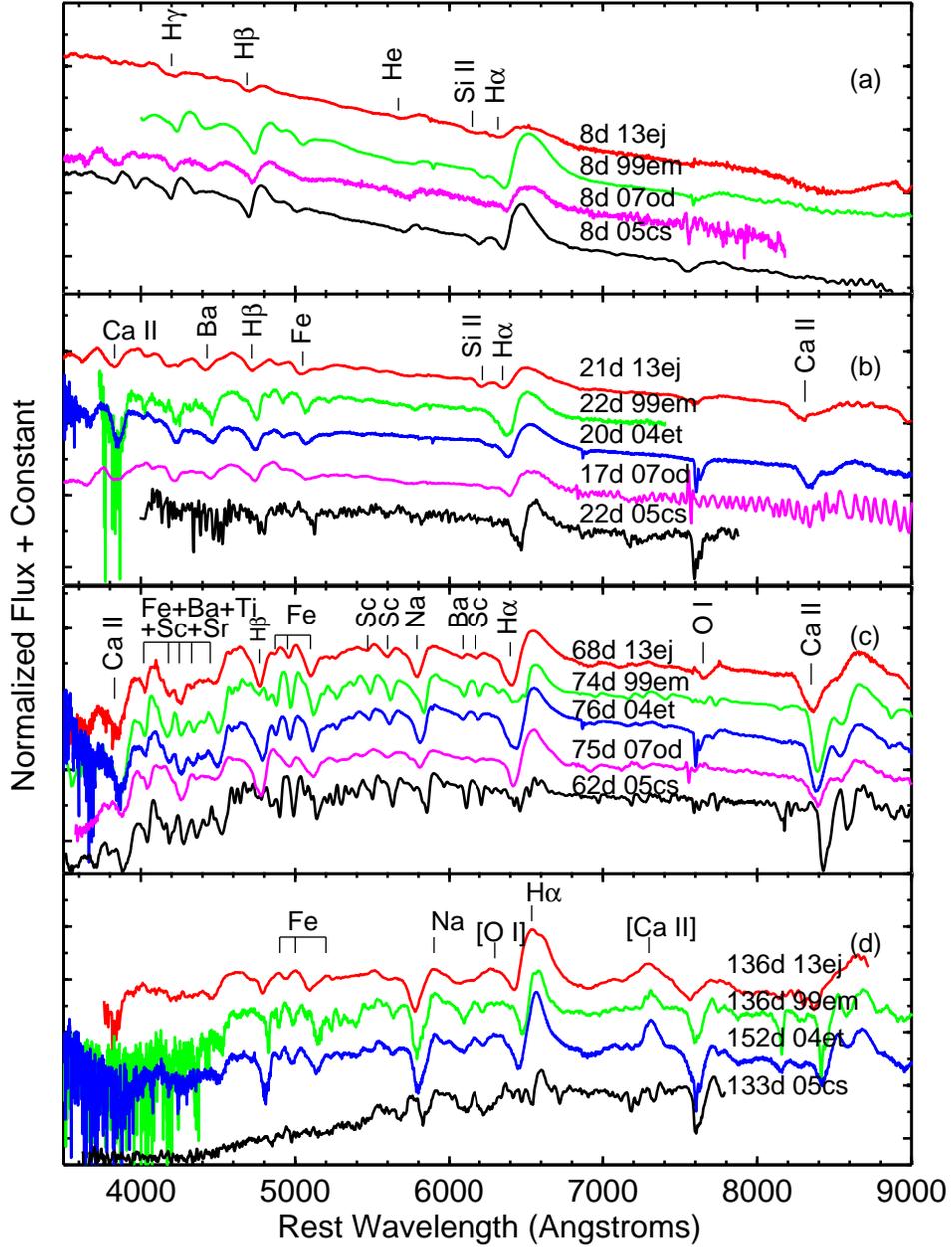}
 \caption{Spectral comparison of SN 2013ej with SNe 1999em, 2004et, 2005cs and 2007od at $\sim$ 1 week, 3 weeks, 2 months, and 5 months after explosion, respectively.}
 \label{fig:speccompare}
 \end{figure}

Figure \ref{fig:speccompare} shows the spectral comparison of SN 2013ej with some representative SNe IIP at t $\sim$ 1 week, $\sim$ 3 weeks, $\sim$ 2 months, and $\sim$ 5 months after explosion. The comparison sample include SNe 1999em \citep{Leonard2002}, 2004et \citep{Sahu2006}, 2005cs \citep{Pastorello2009}, and 2007od \citep{Inserra2011}. All of the spectra have been corrected for redshift and reddening.

At t $\sim$ +8 days, the spectrum is characterized by broad P-Cygni profiles of H Balmer lines superimposed on the blue continuum (see Fig. \ref{fig:speccompare}a). Note that the hydrogen lines of SN 2013ej are less pronounced when compared to the comparison sample. The weak absorption seen near 5700\AA\ in the t=+8 day spectrum is likely due to the He I 5876 feature (see also the spectra at t $\sim$ +10, +11, and +17 days shown in Figure \ref{fig:spectra}), with the peak near 5800\AA\ corresponding to the P-Cygni emission component. While other He I lines can be hardly identified in the spectrum. The absorption at about 6200 \AA~ can be attributed to \ion{Si}{2} $\lambda$6355 rather than high velocity feature of H$\alpha$ for the lack of a similar absorption feature on the blueside of H$\beta$ (\citealt{Valenti2014}; see also the right panel of Figure \ref{fig:Hab}). By about 3 weeks from the explosion, the continuum spectrum flattens as a result of the dramatic decline of the photospheric temperature. At this time, the P-Cygni profiles of H lines are well developed. The possible feature due to He I 5876 becomes barely visible in the t=+21 day spectrum but a few more metal lines such as Na I, Ca II, Ba II $\lambda$4554, and Fe II $\lambda$5169 start to appear in the spectrum (see Fig. \ref{fig:speccompare}b).

By t $\sim$ 2 months after the explosion, the supernova enters the plateau phase. The spectrum is dominated by hydrogen lines and some metal lines such as Fe~II (4500, 4924, 5018, and 5169), Sc~II (5527, 5672, and 6305), Ba II (6142), and Ca~II H\&K and IR triplet, as shown in Fig. \ref{fig:speccompare}c. During the plateau phase (e.g., from t $\sim$ 30 to t $\sim$ 80 days), the spectra show little change except that the absorption troughs of the above features become gradually stronger.

Fig. \ref{fig:speccompare}d shows the comparison at t $\sim$ 5 months. At such a late phase, the absorption features generally become shallower while the emission components grow progressively stronger. In particular, the narrow forbidden emission lines [O~I] 6300, 6364\AA ~and [Ca~II] 7291, 7324\AA ~start to emerge in the spectra.

The spectra of SN 2013ej are somewhat similar to the luminous SNe IIP such as SN 2004et and SN 2007od, but differences do exist between them. For example, SN 2013ej shows a stronger O I absorption at early phases relative to the above two comparison SNe. A stronger O I absorption perhaps indicates a smaller Ca/O ratio for the progenitor of SN 2013ej, which in turn points to a larger progenitor mass because the Ca/O ratio decreases with the progenitor mass \citep{HouckFransson}.  Moreover, SN 2013ej does not seem to show the boxy H$_{\alpha}$ and H$_{\beta}$ profiles (which are formed perhaps due to weak CSM interaction) as seen in SN 2007od , despite that they both have broad line profiles.

\subsection{Ejecta Velocity}
\begin{figure}
  \centering
   \includegraphics{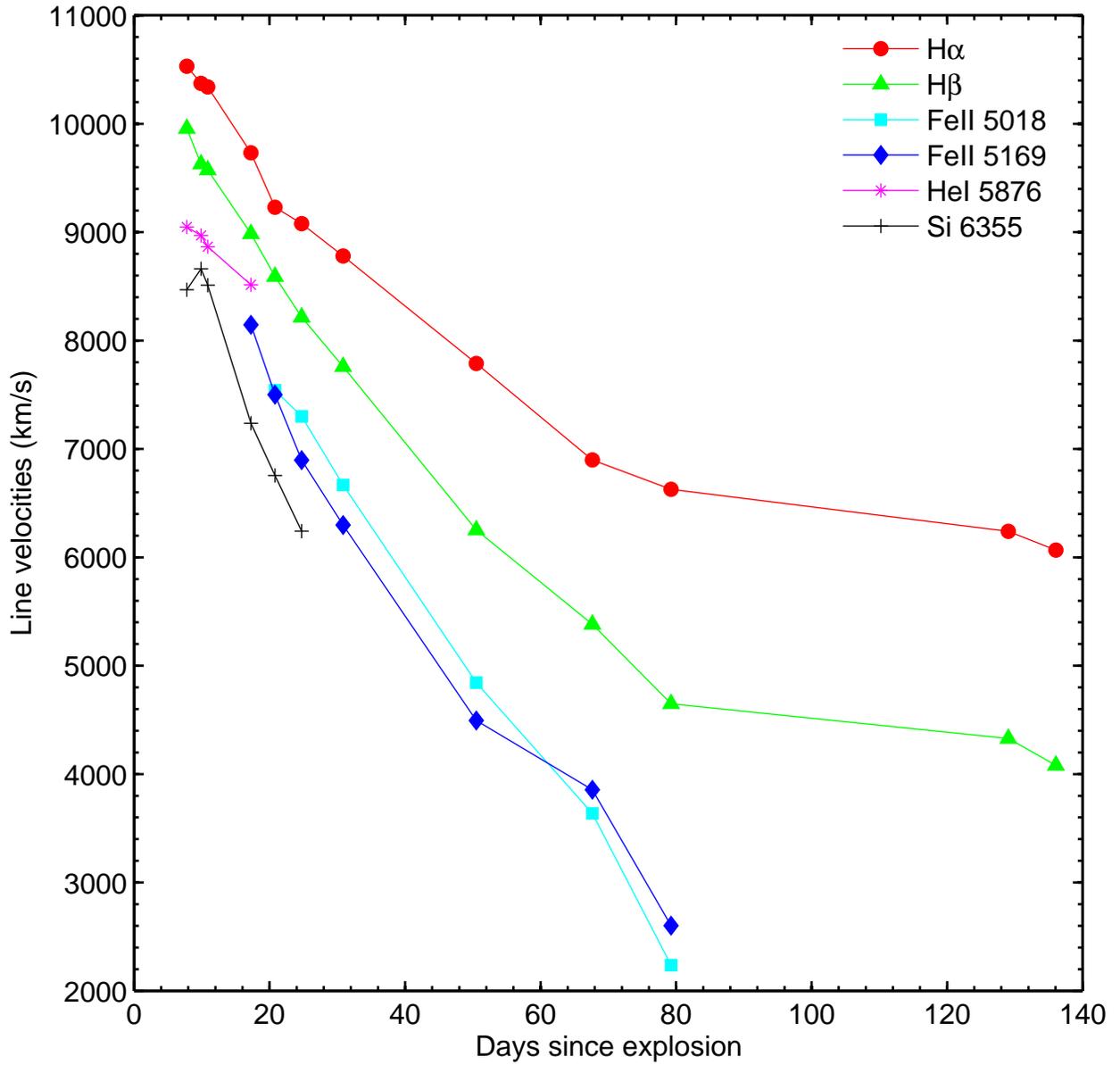}
  \caption{Evolution of the ejecta velocities of SN 2013ej measured from the absorption minima of different spectral lines. }
 \label{fig:vel}
  \end{figure}

The ejecta velocities can be estimated by performing a Gaussian fit to the absorption minima of some unblended lines in the spectra. The velocities inferred from absorption minima of H$\alpha$, H$\beta$, Si II $\lambda$6355, He I $\lambda$5876 and Fe II $\lambda$5018, 5169 are shown in Figure \ref{fig:vel}. The typical uncertainties in the measured velocities are 20--350 km/s, depending on the resolution and the signal to noise (S/N) ratio of the spectra. After t $\sim$ 80 days, we did not attempt to estimate the velocity of Fe II lines because they do not show well-defined Gaussian profiles perhaps due to contaminations by other unknown lines. It can be seen that the expansion velocities of different species decrease with time approximately in a power-law fashion. After t $\sim$ 80 days, the velocities of hydrogen lines decrease continuously but at a much slower pace. Note that the velocities derived from H$\alpha$ and H$\beta$ are always higher than the metal lines, and this can be naturally explained by that these lines are formed at the top of photosphere and have lower optical depths.

\begin{figure}
 \centering
 \includegraphics{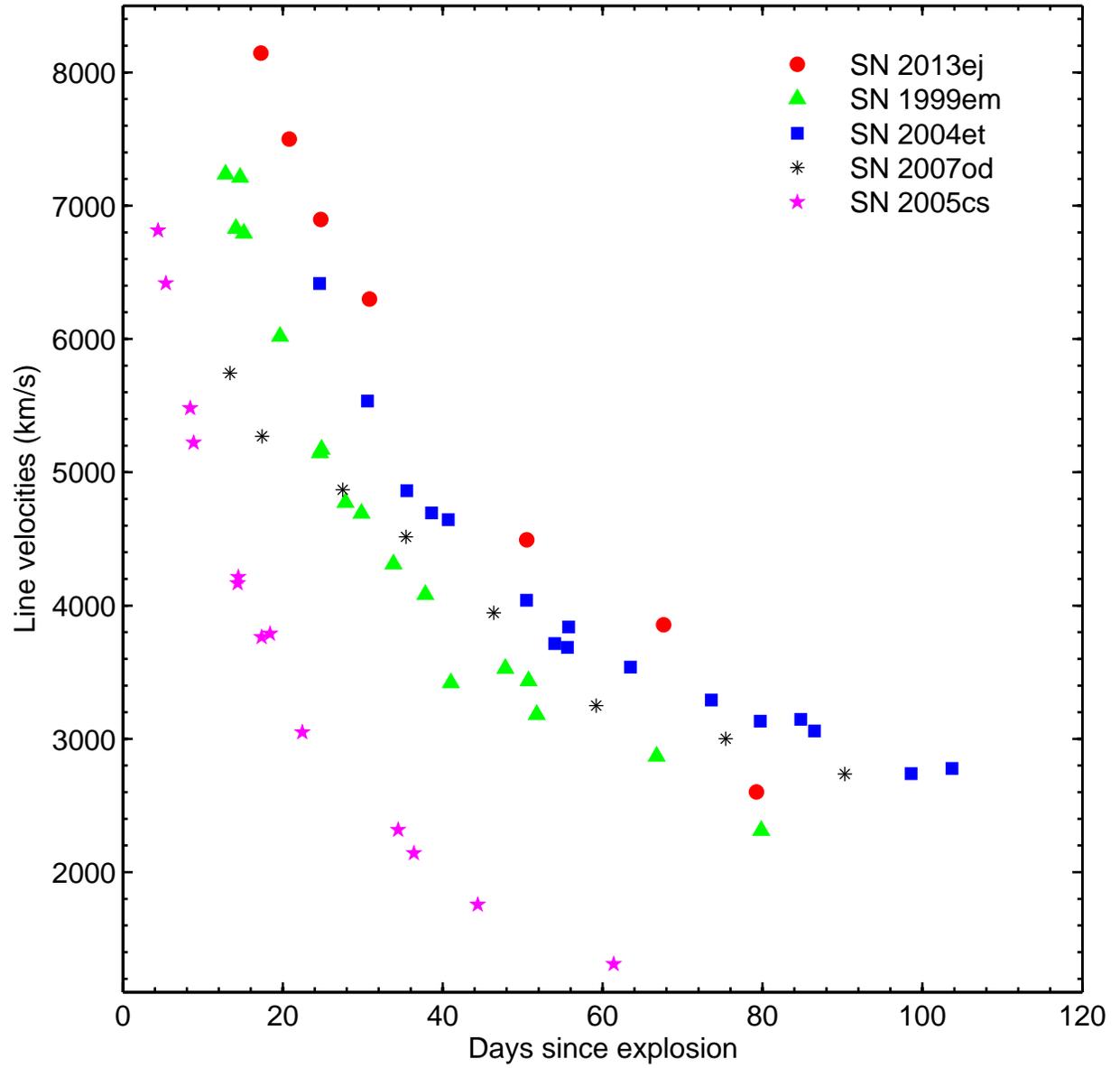}
 \caption{Evolution of photospheric velocity of SN 2013ej inferred from Fe~II$\lambda$5169, compared with those of some well-studied SNe IIP \citep{Takats2012} and SN 2007od \citep{Inserra2011}.}
 \label{fig:velcomp}
\end{figure}

\begin{figure}
 \centering
 \includegraphics{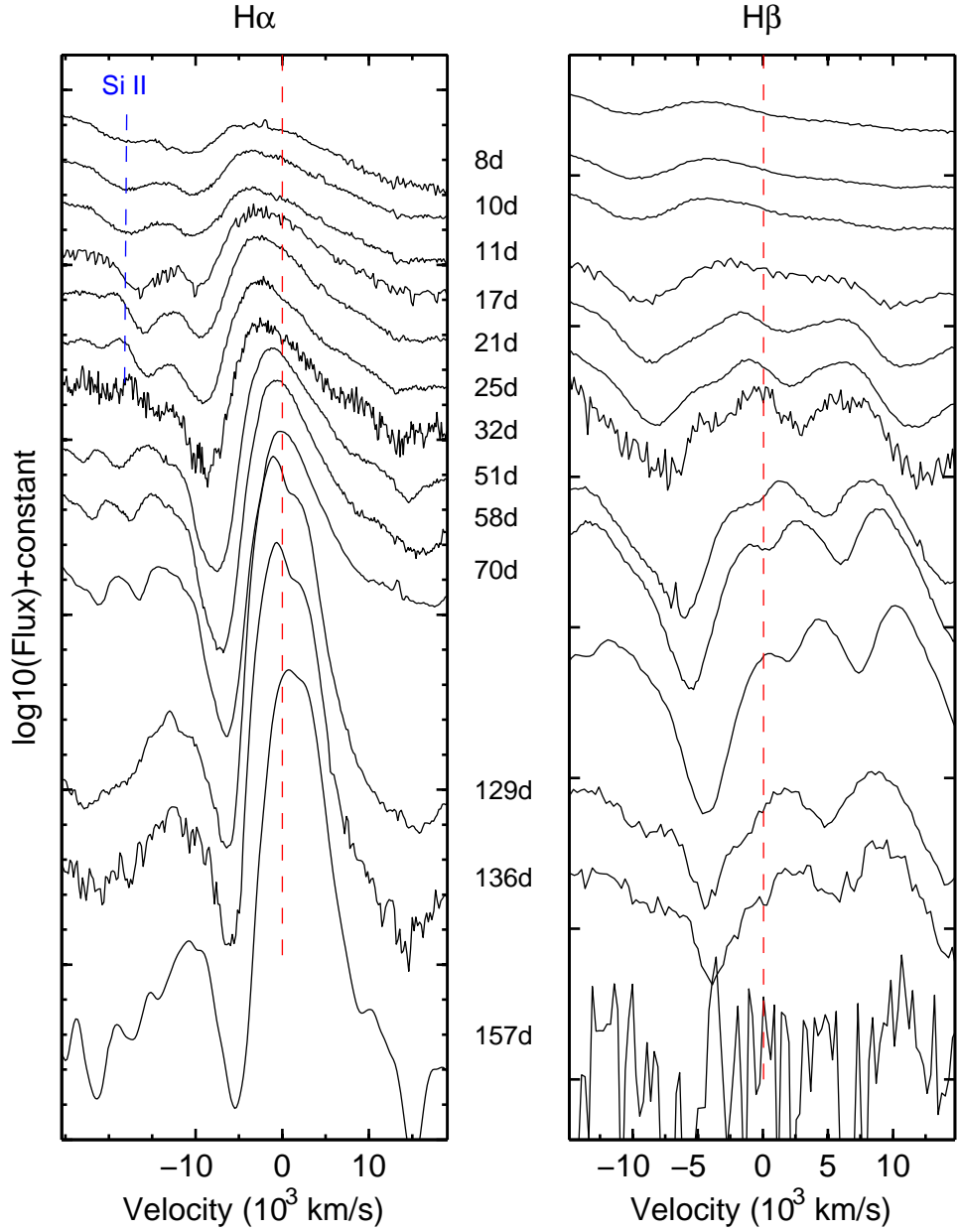}
 \caption{Evolutions of H$\alpha$ 6563 and H$\beta$ 4861 in SN 2013ej. The absorption at about 6200 \AA~, identified as \ion{Si}{2} $\lambda$6355, is marked with a blue dashed line in the left panel.} 
 \label{fig:Hab}
 \end{figure}

Fe~II $\lambda$5169 has long been regarded as a good indicator of the photospheric velocity for SNe IIP \citep{Hamuy2001}, as it is formed close to the photosphere and suffers less blending from other lines in the earlier phases. Figure \ref{fig:velcomp} shows the photospheric velocity of SN 2013ej as measured from Fe II $\lambda$5169, together with those of the comparison sample of SNe IIP. One can see that SN 2013ej has a a higher photospheric velocity relative to the comparison sample at all phases. At t$\sim$ 50 days, the Fe~II velocity is measured to be $\sim$ 4600 km s$^{-1}$ for SN 2013ej, while it is $\sim$3800 km s$^{-1}$ and only $\sim$3200 km s$^{-1}$ for SN 2007od and SN 1999em, respectively. The larger expansion velocity seen in SN 2013ej is generally consistent with its higher luminosity, following the luminosity-velocity correlation proposed by Hamuy et al. (2003). SN 2007od seems to be an outlier of this correlation, and its higher luminosity is likely caused by the ejecta-CSM interaction (Inserra et al. 2011).

\subsection{Asymmetric Line Profiles in Nebular Spectra}
The emission-line profiles in the nebular spectra of SNe are powerful probes of the geometric structure of the ejecta in the explosion. The H$\alpha$ is the strongest line in the nebular phase, and can be used for studies of this kind. The evolution of H$\alpha$ and H$\beta$ line profiles of SN 2013ej is shown in Figure \ref{fig:Hab}. While in the early phase H$\alpha$ was characterized by a symmetric P-Cygni profile, a double peaked emission appears in the t $\sim$ 129 days spectrum, and this asymmetric feature is also visible in the t $\sim$ 136 days and t $\sim$ 157 days spectra. It is not clear whether the double-peak feature exists in H$\beta$ because it is relatively weak and may suffer from line blending. This asymmetric feature of H$\alpha$ was similarly seen before in SN 1987A \citep{Utrobin1995} and SN 2004dj \citep{Chugai2005}, and can be also seen in our comparison sample SN 1999em and SN 2007od at comparable phases but not in the low-luminosity object SN 2005cs, as shown in Figure \ref{fig:speccompare}d. One notable feature in SN 2013ej is that its line profile of H$\alpha$ seems to be much broader than that of the comparison sample, consistent with that its explosion ejecta has a higher expansion velocity. In SN 2013ej, the asymmetric feature can be decomposed into a dominant blue peak shifted by $\sim$ $-$1200 km s$^{-1}$ and a red peak shifted by $\sim$ +1000 km s$^{-1}$. This phenomenon may indicate an asymmetric (bipolar) ejection of $^{56}$Ni in a spherically symmetric envelope, as suggested by \cite{Chugai2006} in study of the double-peak feature of SN 2004dj. The stronger blueside peak can be explained with a $^{56}$Ni distribution that is skewed towards the observer, while the red peak can be attributed to greater scattering and absorptions for photons traveling from the far side of the ejecta.

\section{Discussions} \label{sect:discuss}

\subsection{The Bolometric Light Curve and the Nickel Mass}
\label{subsect:Nimass}
Using the UV, optical, and NIR data presented in previous sections, we can compute the $UVOIR$ bolometric light curve of SN 2013ej. The bolometric luminosity is derived by correcting the observed magnitudes for extinctions, converting the magnitudes into fluxes at the effective wavelength, and finally integrating the resulting spectral energy distribution (SED) over wavelengths. The integrated flux is then converted to luminosity by adopting the distance of 9.6 Mpc (see \S \ref{subsect:distance}). The flux was calculated for epochs when the V-band magnitudes were obtained; and the data in other bands were interpolated from the adjacent color curves using a low-order polynomial fit whenever necessary.

\begin{figure}
 \centering
 \includegraphics{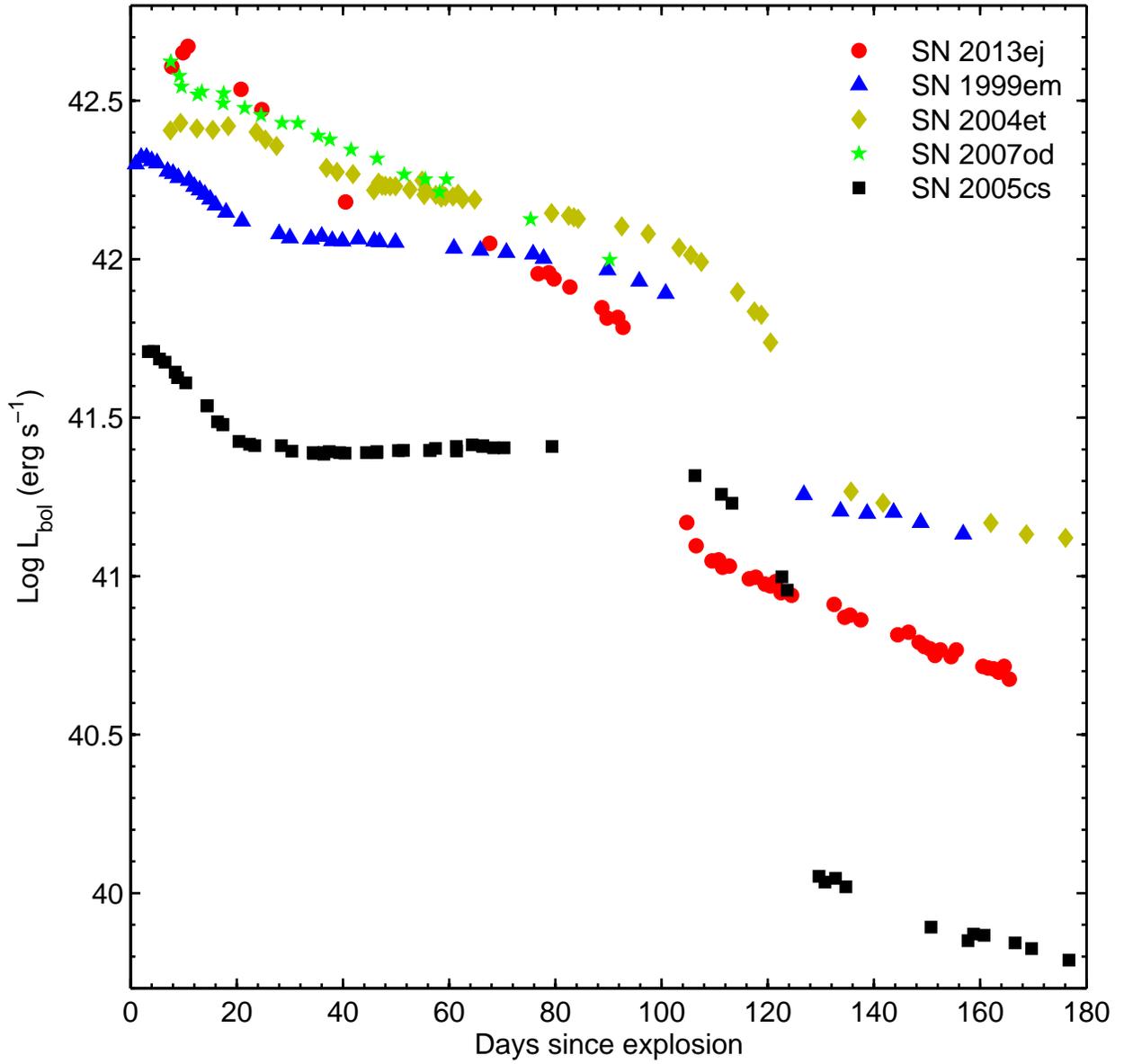}
 \caption{The $UBVRI$ pseudo-bolometric light curve of SN 2013ej, compared with those of other well-studied SNe IIP.}
 \label{fig:bolo}
\end{figure}

As the UV and NIR data were not available for some of our sample, we present in Figure \ref{fig:bolo} a comparison of the $UBVRI$ pseudo-bolometric light curves of SN 2013ej with those of other SNe IIP. As shown in the plot, SN 2013ej has a higher peak luminosity and a shorter plateau phase than the comparison SNe IIP. In the nebular phase (i.e., at t $\sim$ 120--170 days), its luminosity shows a faster decline, with a magnitude decline rate of 1.52 ${\rm mag}\,(100{\rm d})^{-1}$. As a comparison, the magnitude decline rate per 100 day is found to be 0.92 mag for SN 1999em, 0.94 mag for SN 2005cs, 0.91 mag for SN 2004et, and 1.05 mag for SN 2007od, respectively. This may suggest that the ejecta of SN 2013ej became optically thinner than other SNe IIP at comparable phases. A smaller opacity can be also inferred for SN 2013ej from its higher luminosity and shorter plateau phase according to the semi-analytical light curve model presented by Nagy et al. (2014). For a type II supernova, the opacity of the photosphere is primarily determined by absorptions of neutral hydrogen in the earlier phase and negatively-ionized hydrogen in the nebular phase (due to its lower ionization energy, i.e., 0.754 ev). It is thus reasonable to speculate that the progenitor SN 2013ej has a lower-mass envelope of hydrogen.

The $^{56}$Ni mass ejected in the nebular phase by SN 2013ej can be derived by comparing the $UVOIR$ bolometric light curve to that of SN 1987A, assuming a similar $\gamma$-ray deposition fraction:

\begin{equation}
 M(^{56}Ni)_{13ej}=M(^{56}Ni)_{87A} \times \frac{L_{13ej}}{L_{87A}} M_{\odot}
\end{equation}

where $M(^{56}Ni)_{87A} =0.075\pm0.005 M_{\odot}$ is the mass of $^{56}$Ni ejected by SN 1987A \citep{Arnett1996}, and $L_{87A}$ is the $UVOIR$ bolometric luminosity at a comparable epoch. The comparison with the tail evolution of SN 1987A during the phase from t $\sim$ 140--170 days gives M$(^{56}Ni)_{13ej}$ $\sim$ 0.02 $\pm$ 0.01 M$_{\odot}$.

\subsection{Progenitor of SN 2013ej}  \label{subsect:progenitor}
The progenitor and explosion properties of SNe IIP can be studied by direct progenitor detection in the pre-explosion images. Based on the archival HST ACS images, \cite{Fraser2014} suggested the progenitor of SN 2013ej to be an M-type supergiant with the mass of 8--16 M$_{\odot}$ and the luminosity of $log L/L_{\odot}$ = 4.46--4.85 dex. Using the temperature evolution of the photosphere at early phases, \cite{Valenti2014} set an interesting constraint on the radius of the progenitor, i.e. 400--600 R$_{\odot}$, which is comparable to that of a typical RSG.

Like the analysis done for some SNe IIP (e.g., SNe 2007od, 2009bw, and 2009E; see \citealt{Inserra2011, Inserra2012, Pastorello2012}), we attempt to derive the progenitor parameters for SN 2013ej using a two-step modeling procedure. The first step is solving the energy balance equation for a spherically symmetric envelope of constant density in homologous expansion through a semi-analytic code \citep{Zampieri2003}. The second step involves solving the equations of relativistic radiation hydrodynamics, which is to compute the physical properties of the ejecta and the evolution of the main observables (up to the nebular phase)\citep{Pumo2010, Pumo2011}. Compared to the semi-analytic model, the general-relativistic, radiation-hydrodynamics model is more realistic but time consuming. In our analysis, the semi-analytic code is first used to carry out a preparatory study aiming at individuating the parameter space describing the SN progenitor at the explosion. The results from such a study are then used as inputs for the model calculations performed with the radiation-hydrodynamics code \citep{Pumo2010, Pumo2011}.

\begin{figure}
\centering
\includegraphics{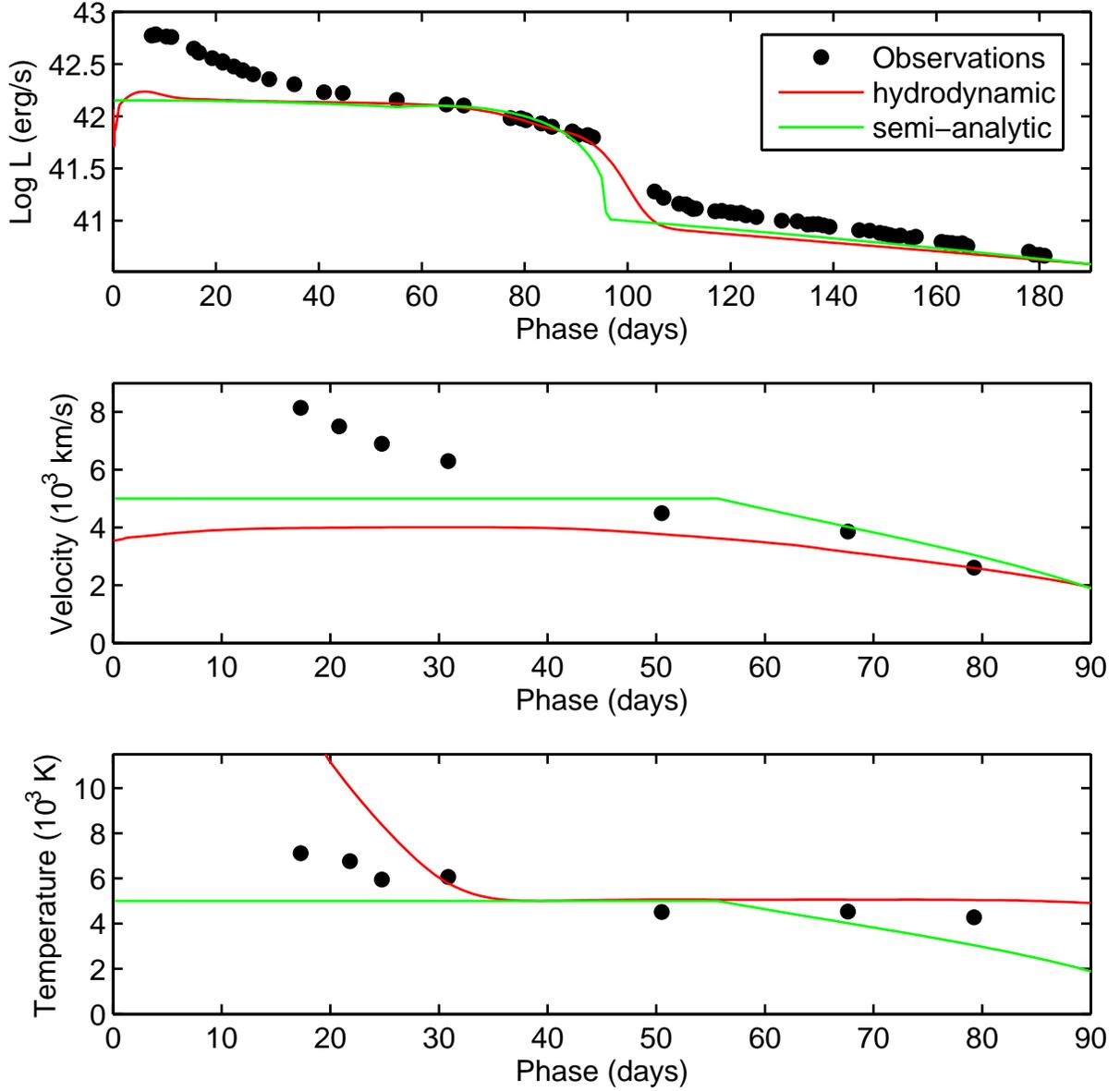}
\caption{Comparison of the evolution of the main observables (bolometric luminosity, photospheric velocity, and photospheric temperature) of SN 2013ej with the best-fit relativistic, radiation-hydrodynamic model (total energy 0.7 foe, initial radius $4.2 \times 10^{13}$ cm, and ejected mass 10.6 M$_{\odot}$). The best-fit semi-analytic model is also shown for comparison (total energy $\sim$2.1 foe, initial radius $\sim$1.6$\times$10$^{13}$ cm, and ejected mass $\sim$10.4 M$_{\odot}$). The photospheric velocities are derived from the minimum absorption of Fe II $\lambda$5169.}.
\label{fig:model}
\end{figure}

The photospheric temperatures were derived by applying blackbody fit to the continuums of the observed spectra (which have been corrected for redshift and reddening and adjusted in shape according to the photometry) up to t$\sim$80 days after explosion. For the spectra taken after that, the fit to the continuum becomes difficult due to the increasing line blanketing and appearances of strong emission lines. In Figure \ref{fig:model}, we present the comparison of the observed values with the best-fit model calculations. The agreement between our modelings and the observations is reasonably good, except for the velocity evolution at early phase (which happens also for other SNe IIP). The discrepancy in the velocity evolution can be attributed to the fact that the radial density profile is not perfectly reproduced in the outermost shells of the ejecta formed after shock breakout. For this reason, we exclude the early data from the fit. Overall, the hydro-dynamical model fits the data better than the semi-analytic model, especially for the bolometric light curve and the temperature.

Assuming a $^{56}$Ni mass of 0.020 M$_{\odot}$ from the observed light curves, the best-fit model computed using the semi-analytic and radiation-hydrodynamical models yields a total (kinetic plus thermal) energy of 0.7--2.1$\times$ 10$^{51}$ erg, an initial radius of 1.6--4.2$\times$10$^{13}$ cm ($\sim$ 230--600 R$_{\odot}$), and an envelope mass of 10.4--10.6 M$_{\odot}$. Taking into account approximately 1.5--2.0 M$_{\odot}$ for the remnant neutron star, we obtain a range of 12--13 M$_{\odot}$ for the mass of the progenitor immediately before explosion, which is in agreement with the direct observational estimate from \cite{Fraser2014}.

\section{Conclusions}  \label{sect:sum}
In this paper, we present extensive $UVOIR$ observations of SN 2013ej. The photometric observations are presented at 47 epochs covering from +8 days to +185 days after the explosion, and the low-resolution spectroscopic observations are presented at 13 epochs spanning from +8 days to +157 days.

The analyses of the photometric and spectroscopic observations of SN 2013ej suggest that it belongs to the subclass of SNe IIP but shows differences in some aspects. Compared to normal SNe IIP such as SN 1999em, SN 2013ej is more luminous around the peak and shows faster decline rates after that, leading to a shorter plateau phase of about 50 days. The spectral evolution of SN 2013ej shows close resemblance to that of SN 2004et and SN 2007od at comparable phases, but the former has higher expansion velocities and broader line profiles. At nebular phase, the emission of H$\alpha$ displayed a pronounced asymmetric structure, which may be due to the asymmetric ejection of $^{56}$Ni in a spherically symmetric but thinner envelope.

A nickel mass of 0.02$\pm$0.01 M$_{\odot}$ can be obtained for SN 2013ej using the tail bolometric light curve. With the hydrodynamical model and input parameters from our observations, we are able to reproduce some physical properties for the progenitor of SN 2013ej. The simulation yields a total explosion energy of 0.7 $\times$ 10$^{51}$ erg, a radius of the progenitor as $\sim$ 600 R$_{\odot}$, and an ejected mass of $\sim$ 10.6 M$_{\odot}$, consistent with an RSG with a mass of 12--13 M$_{\odot}$ before explosion. The observational evidence presented for SN 2013ej, together with significant polarization measured at early phase, suggest that its progenitor may lose a considerable amount of hydrogen before explosion.

\section*{Acknowledgments}
We thank all the staffs at the Li-Jiang Observatory, Yunnan Astronomical Observatory of China and Xinglong Station, National Astronomical Observatory of China for the observations and technological support. We acknowledge the use of public data from the $Swift$ data archive. This work is also based on observations made with ESO Telescopes at the La Silla Paranal Observatory under programme IDs 188.D-2003 and 191.D-0935. The work of X. Wang is supported by the Major State Basic Research Development Program (2013CB834903), the National Natural Science Foundation of China (NSFC grants 11178003 and 11325313), the Foundation of Tsinghua University (2011Z02170), and the Strategic Priority Research Program ``The Emergence of Cosmological Structures" of the Chinese Academy of Sciences (grant No. XDB09000000). J.-J. Zhang is supported by the NSFC (grants 11403096), and T.-M. Zhang is supported by the NSFC (grants 11203034).

\clearpage

\clearpage

   \begin{table}
   \caption{Photometric Standard Stars in the SN 2013ej Field.}  \label{tab:localstar}
      \centering

   \begin{tabular}{cccccccc}
   \hline
       Star& $\alpha_{\rm J2000}$&       $\delta_{\rm J2000}$& $U$ & $B$&  $V$&  $R$&  $I$\\
       ID&          (h m s)      &(\degr\,\arcmin\,\arcsec)  &(mag)&(mag)&(mag)&(mag)&(mag)\\
       \hline
       1 &  01:37:09.07 & 15:48:00.0 & 17.02(02) & 16.17(06) &  15.18(04)  & 14.60(04) &  14.06(07)\\
       2 &  01:36:58.61 & 15:47:46.3 & 12.84(03) & 12.95(02) &  12.51(01)  & 12.26(02) &  11.98(01)\\
       3 &  01:36:59.63 & 15:46:47.5 & 17.13(02) & 16.06(06) &  14.97(04)  & 14.34(04) &  13.79(08)\\
       4 &  01:36:57.57 & 15:46:20.1 & 13.93(03) & 13.75(03) &  13.12(02)  & 12.74(01) &  12.42(02) \\
       5 &  01:36:27.39 & 15:46:29.5 & 14.73(03) & 14.73(09) &  14.07(21)  & 13.57(18) &  13.04(18)\\
       6 &  01:36:32.10 & 15:45:07.3 & 15.28(05) & 14.15(03) &  13.03(02)  & 12.29(01) &  11.64(01)\\
       7 &  01:37:09.09 & 15:41:20.0 & 18.70(04) & 17.81(06) &  16.75(03)  & 16.14(04) &  15.58(08)\\
       8 &  01:37:03.49 & 15:41:38.7 & 17.19(02) & 17.08(04) &  16.40(03)  & 16.00(03) &  15.60(06)\\
       9 &  01:36:52.78 & 15:40:38.6 & 14.81(02) & 14.24(04) &  14.38(03)  & 14.34(03) &  13.92(06)\\
      10 &  01:36:47.02 & 15:41:19.7 & 18.17(03) & 17.64(05) &  16.80(03)  & 16.30(04) &  15.81(07)\\
   \hline
   \end{tabular}
   \end{table}

     \begin{table}
     \fontsize{2.7mm}{2.7mm}\selectfont
         \caption{Photometric evolution of SN 2013ej.} \label{tab:photsn}
    \begin{tabular}{ccccccccc}
     \hline
  UT Date   & JD      &Phase$^{a}$&  $U$   &$B$   &$V$  &$R$    &$I$   &Telescope$^{b}$ \\
  (yy/mm/dd)&2450000+ &(day)      &(mag)   &(mag) &(mag) &(mag) &(mag)& \\
  \hline
2013-07-31 & 6504.28 &   7.28 & \nodata   & 12.62(02) & 12.58(03) & 12.66(04) & 12.572(03)  & LJT \\
2013-08-02 & 6506.38 &   9.38 & \nodata   & 12.58(02) & 12.49(04) & 12.44(08) & 12.310(06)  & LJT \\
2013-08-03 & 6507.32 &  10.32 & \nodata   & 12.54(03) & 12.48(12) & 12.28(09) & 12.211(08)  & LJT \\
2013-08-13 & 6517.31 &  20.31 & \nodata   & 13.05(04) & 12.52(04) & 12.31(09) & 12.203(09)  & LJT \\
2013-08-17 & 6521.24 &  24.24 & \nodata   & 13.33(05) & 12.61(05) & 12.45(07) & 12.183(09)  & LJT \\
2013-09-12 & 6537.01 &  40.01 & 15.53(03) & 14.55(01) & 13.22(06) & 12.88(07) & 12.381(06)  & LJT \\
2013-09-29 & 6564.14 &  67.14 & 16.40(08) & 14.91(09) & 13.57(07) & 13.20(06) & 12.599(07)  & LJT \\
2013-10-08 & 6573.25 &  76.25 & 16.55(07) & 15.14(05) & 13.74(05) & 13.33(05) & 13.124(07) & TNT \\
2013-10-10 & 6575.25 &  78.25 & 16.50(05) & 15.23(04) & 13.81(03) & 13.28(04) & 13.015(03) & TNT \\
2013-10-11 & 6576.25 &  79.25 & 16.79(05) & 15.32(05) & 13.81(03) & 13.35(05) & 13.053(06) & TNT \\
2013-10-14 & 6579.25 &  82.25 & 16.85(05) & 15.39(04) & 13.91(04) & 13.40(05) & 13.088(06) & TNT \\
2013-10-20 & 6585.25 &  88.25 & 16.71(15) & 15.56(05) & 14.05(04) & 13.60(05) & 13.246(02) & TNT \\
2013-10-21 & 6586.25 &  89.25 & 16.83(18) & 15.67(07) & 14.09(05) & 13.66(04) & 13.407(04) & TNT \\
2013-10-23 & 6588.25 &  91.25 & \nodata   & 15.69(05) & 14.19(03) & 13.59(03) & 13.328(02) & TNT \\
2013-10-24 & 6589.25 &  92.25 & \nodata   & 15.77(04) & 14.27(03) & 13.71(03) & 13.303(04) & TNT \\
2013-11-05 & 6601.25 & 104.25 & 18.20(13) & 17.47(04) & 16.02(03) & 15.15(04) & 14.797(03) & TNT \\
2013-11-07 & 6603.00 & 106.00 & 18.43(10) & 17.57(03) & 16.20(03) & 15.36(03) & 14.969(03) & TNT \\
2013-11-10 & 6606.00 & 109.00 & \nodata   & 17.76(05) & 16.39(02) & 15.42(03) & 15.093(04) & TNT \\
2013-11-11 & 6607.25 & 110.25 & 18.44(15) & 17.79(04) & 16.40(03) & 15.43(04) & 15.049(04) & TNT \\
2013-11-12 & 6608.00 & 111.00 & \nodata   & 17.62(08) & 16.45(04) & 15.52(06) & 15.157(04) & TNT \\
2013-11-13 & 6609.25 & 112.25 & \nodata   & \nodata   & 16.55(08) & 15.45(06) & \nodata   & TNT \\
2013-11-17 & 6613.00 & 116.00 & \nodata   & \nodata   & 16.60(08) & 15.60(05) & 15.26(05) & TNT \\
2013-11-18 & 6614.25 & 117.25 & \nodata   & 17.65(16) & 16.60(05) & 15.58(05) & 15.23(04) & TNT \\
2013-11-20 & 6616.00 & 119.00 & \nodata   & 17.93(08) & 16.61(04) & 15.64(06) & 15.20(03) & TNT \\
2013-11-21 & 6617.00 & 120.00 & 18.84(15) & 17.92(06) & 16.58(02) & 15.66(02) & 15.26(04) & TNT \\
2013-11-22 & 6618.00 & 121.00 & \nodata   & 17.93(05) & 16.58(04) & 15.57(03) & 15.25(02) & TNT \\
2013-11-23 & 6619.00 & 122.00 & 18.87(11) & 17.97(05) & 16.64(03) & 15.69(04) & 15.32(03) & TNT \\
2013-11-25 & 6621.00 & 124.00 & 18.83(11) & 18.00(05) & 16.64(03) & 15.69(04) & 15.38(03) & TNT \\
2013-12-03 & 6629.00 & 132.00 & 18.81(10) & 18.06(05) & 16.72(06) & 15.76(03) & 15.47(03) & TNT \\
2013-12-05 & 6631.00 & 134.00 & \nodata   & 18.11(05) & 16.81(04) & 15.79(04) & 15.64(04) & TNT \\
2013-12-06 & 6632.00 & 135.00 & \nodata   & 18.12(04) & 16.81(04) & 15.78(02) & 15.58(04) & TNT \\
2013-12-07 & 6633.00 & 136.00 & \nodata   & \nodata   & 16.79(04) & 15.79(04) & \nodata   & TNT \\
2013-12-08 & 6634.00 & 137.00 & \nodata   & 18.00(05) & 16.90(03) & 15.86(03) & 15.57(03) & TNT \\
2013-12-15 & 6641.00 & 144.00 & \nodata   & 18.22(14) & 17.01(06) & 15.97(05) & 15.66(02) & TNT \\
2013-12-17 & 6643.00 & 146.00 & \nodata   & 18.17(17) & 17.07(06) & 15.91(04) & 15.67(02) & TNT \\
2013-12-19 & 6645.00 & 148.00 & \nodata   & 18.54(13) & 17.04(04) & 15.98(03) & 15.71(02) & TNT \\
2013-12-20 & 6646.00 & 149.00 & \nodata   & 18.46(11) & 17.16(04) & 16.04(03) & 15.69(03) & TNT \\
2013-12-21 & 6647.00 & 150.00 & \nodata   & 18.25(07) & 17.12(04) & 16.09(04) & 15.79(03) & TNT \\
2013-12-22 & 6648.00 & 151.00 & \nodata   & 18.38(05) & 17.10(03) & 16.15(04) & 15.83(02) & TNT \\
2013-12-23 & 6649.00 & 152.00 & \nodata   & 18.34(06) & 17.03(05) & 16.13(06) & 15.78(03) & TNT \\
2013-12-25 & 6651.00 & 154.00 & \nodata   & 18.44(07) & 17.13(03) & 16.12(04) & 15.89(04) & TNT \\
2013-12-26 & 6652.00 & 155.00 & \nodata   & 18.34(07) & 17.11(03) & 16.08(03) & 15.81(03) & TNT \\
2013-12-31 & 6657.00 & 160.00 & \nodata   & 18.41(06) & 17.20(05) & 16.23(04) & 15.97(03) & TNT \\
2014-01-01 & 6658.00 & 161.00 & \nodata   & 18.35(06) & 17.22(06) & 16.24(04) & 16.01(04) & TNT \\
2014-01-02 & 6659.00 & 162.00 & \nodata   & 18.42(12) & 17.23(07) & 16.23(07) & 16.01(05) & TNT \\
2014-01-03 & 6660.00 & 163.00 & \nodata   & 18.61(07) & 17.26(04) & 16.23(05) & 16.01(04) & TNT \\
2014-01-04 & 6661.00 & 164.00 & \nodata   & 18.41(13) & 17.21(06) & 16.19(05) & 16.03(05) & TNT \\
2014-01-05 & 6662.00 & 165.00 & \nodata   & 18.55(06) & 17.32(04) & 16.32(03) & 16.05(02) & TNT \\
  \hline
 \end{tabular}
 \tablecomments{Uncertainties (numbers in brackets), in units of 0.01 mag, are 1$\sigma$.}
 \begin{flushleft}
  $^{a}$ With reference to the explosion epoch JD 2456497.0\\
  $^{b}$ LJT : 240-cm Lijiang Telescope, Yunnan, China; TNT : 80-cm Tsinghua-NAOC telescope, Hebei, China\\
 \end{flushleft}
    \end{table}

 \begin{table}
       \fontsize{3mm}{3mm}\selectfont
  \caption{Ultraviolet and Optical photometry of SN 2013ej from the $Swift$ UVOT}    \label{tab:uvot}
   \begin{tabular}{ccccccccc}
   \hline
   UT Date &    JD       & Phase$^{a}$ & $uvw$2       &    $uvm$2   &    $uvw$1  &   $U$ &     $B$    & $V$   \\
   (yy/mm/dd)&  2450000+ & (day)      & (mag)        &    (mag)    &    (mag)   &   (mag)&   (mag)   & (mag) \\
  \hline
2013-07-30&    6504.46&    7.46&      12.37(04)&    12.02(04)&    11.71(04)&   \nodata  &    \nodata  &    12.69(04)\\
2013-07-31&    6505.32&    8.32&      12.58(03)&    12.20(04)&    11.81(03)&   \nodata  &    \nodata  &    12.61(04)\\
2013-08-02&    6507.55&   10.55&      13.05(04)&    12.70(04)&    \nodata  &   11.68(03)&    12.62(03)&    \nodata  \\
2013-08-04&    6509.34&   12.34&      13.38(04)&    13.16(05)&    \nodata  &   11.81(03)&    12.61(03)&    \nodata  \\
2013-08-07&    6511.95&   14.95&      13.93(05)&    \nodata  &    12.95(04)&   \nodata  &    \nodata  &    \nodata  \\
2013-08-08&    6512.68&   15.68&      14.12(05)&    13.97(06)&    13.13(04)&   12.20(03)&    12.75(03)&    12.48(03)\\
2013-08-09&    6513.71&   16.71&      14.39(05)&    14.41(06)&    13.38(04)&   12.33(03)&    12.91(03)&    12.54(03)\\
2013-08-11&    6516.28&   19.28&      15.11(08)&    15.15(08)&    13.91(05)&   12.66(03)&    12.98(03)&    12.58(03)\\
2013-08-13&    6518.33&   21.33&      15.65(08)&    15.96(07)&    14.45(06)&   12.98(03)&    13.11(03)&    12.60(03)\\
2013-08-15&    6520.49&   23.49&      16.21(09)&    \nodata  &    14.91(07)&   13.31(03)&    13.22(03)&    12.57(03)\\
2013-08-17&    6522.15&   25.15&      16.59(10)&    17.11(20)&    15.20(07)&   13.60(03)&    13.29(03)&    12.66(03)\\
2013-08-19&    6524.22&   27.22&      16.82(11)&    17.55(22)&    15.49(08)&   13.96(03)&    13.48(03)&    12.69(03)\\
2013-08-22&    6527.34&   30.34&      17.21(12)&    18.22(17)&    15.88(08)&   14.35(05)&    13.63(03)&    12.89(03)\\
2013-08-27&    6532.24&   35.24&      17.75(15)&    18.57(21)&    16.36(10)&   14.84(06)&    13.92(03)&    12.97(04)\\
2013-09-06&    6541.91&   44.91&      18.13(12)&    19.14(19)&    16.79(08)&   15.57(07)&    14.37(03)&    13.23(04)\\
2013-09-16&    6552.23&   55.23&      18.69(16)&    19.49(24)&    17.29(10)&   16.29(07)&    14.75(04)&    13.47(04)\\
2013-09-26&    6561.94&   64.94&      18.79(17)&    \nodata  &    17.56(12)&   16.59(09)&    14.92(04)&    13.60(04)\\
2013-10-06&    6572.32&   75.32&      19.24(23)&    19.88(33)&    17.92(13)&   17.06(09)&    \nodata  &    \nodata  \\
2013-10-16&    6582.28&   85.28&      19.29(25)&    \nodata  &    18.13(17)&   17.41(12)&    15.46(06)&    14.03(05)\\
2013-10-26&    6592.43&   95.43&      \nodata  &    \nodata  &    18.25(19)&   17.51(17)&    \nodata  &    \nodata  \\
2013-11-13&    6610.18&  113.18&      \nodata  &    \nodata  &    19.52(35)&   19.06(31)&    17.97(21)&    16.64(07)\\
2013-11-20&    6616.86&  119.86&      \nodata  &    \nodata  &    \nodata  &   18.98(21)&    17.89(09)&    16.72(08)\\
2013-11-25&    6622.00&  125.00&      \nodata  &    \nodata  &    \nodata  &   19.16(32)&    \nodata  &    \nodata  \\
2013-11-30&    6626.59&  129.59&      \nodata  &    \nodata  &    19.73(31)&   19.34(28)&    18.16(10)&    16.83(08)\\
2013-12-09&    6636.15&  139.15&      \nodata  &    \nodata  &    19.81(33)&   19.34(27)&    18.20(10)&    16.93(09)\\
\hline
   \end{tabular}
   \tablecomments{Uncertainties (numbers in brackets), in units of 0.01 mag, are 1$\sigma$.}
\begin{flushleft}
  $^{a}$ With reference to the explosion epoch JD 2456497.0\\
   \end{flushleft}
  \end{table}

 \begin{table}
  \caption{$JHK$ Magnitudes of SN 2013ej from NTT+SOFI.}    \label{tab:nir}
   \begin{tabular}{cccccc}
   \hline
  UT Date   & JD      &Phase$^{a}$ &  $J$   &$H$   &$K$  \\
  (yy/mm/dd)&2450000+ &(day)      &(mag)   &(mag) &(mag) \\
\hline
2013-08-17 & 6521.77  & 23.77   & 11.87(30) & 11.97(27) & 11.87(30) \\
2013-08-29 & 6533.70  & 35.70   & 11.99(21) & 11.89(39) & 11.52(21) \\
2013-10-04 & 6569.73  & 71.73   & 12.30(29) & 12.33(31) & 12.08(29) \\
2013-11-03 & 6599.56  &101.56   & 13.83(29) & 13.75(41) & 13.56(24) \\
2013-11-13 & 6609.74  &111.74   & 14.45(18) & 14.24(31) & 14.10(14) \\
2013-12-10 & 6636.53  &138.53   & 15.10(16) & 14.71(39) & 14.24(27) \\
2014-01-25 & 6682.52  &184.52   & 15.95(27) & 15.80(40) & 14.95(31) \\
\hline
   \end{tabular}
   \tablecomments{Uncertainties (numbers in brackets), in units of 0.01 mag, are 1$\sigma$.}
\begin{flushleft}
  $^{a}$ With reference to the explosion epoch JD 2456497.0\\
   \end{flushleft}
\end{table}

\begin{table}
\fontsize{3mm}{3mm}\selectfont
 \caption{Journal of Optical Spectroscopic Observations of SN 2013ej.}  \label{tab:speclog}
 \begin{center}
  \begin{tabular}{lcccccc}
   \hline \hline
   Date        &   JD        &    Phase\tablenotemark{a} & Range      & Resolution  & Exposure &Instrument\\
   (dd/mm/yy)  & (2450000+)  &    (days)                 & (\AA)      & (\AA)       & (s)      &          \\
   \hline
   31/07/2013  &   6504.79   &      8                    & 3400--9000  &  18         &  1200    & Lijiang 2.4-m telescope + YFOSC + gm3   \\
   02/08/2013  &   6506.89   &     10                    & 3400--9000  &  18         &  1200    & Lijiang 2.4-m telescope + YFOSC + gm3   \\
   03/08/2013  &   6507.88   &     11                    & 3400--9100  &  18         &  1200    & Lijiang 2.4-m telescope + YFOSC + gm3   \\
   09/08/2013  &   6514.27   &     17                    & 3400--8900  &  3          &   900    & Xinglong 2.16-m telescope + BFOSC + gm4  \\
   13/08/2013  &   6517.81   &     21                    & 3400--9100  &  18         &   976    & Lijiang 2.4-m telescope + YFOSC + gm3   \\
   17/08/2013  &   6521.75   &     25                    & 3400--9100  &  18         &  1200    & Lijiang 2.4-m telescope + YFOSC + gm3    \\
   24/08/2013  &   6527.85   &     31                    & 3400--9000  &  18         &   263    & Lijiang 2.4-m telescope + YFOSC + gm3    \\
   12/09/2013  &   6547.51   &     51                    & 3300--9100  &  18         &  1200    & Lijiang 2.4-m telescope + YFOSC + gm3    \\
   29/09/2013  &   6564.65   &     68                    & 3300--9100  &  18         &  1800    & Lijiang 2.4-m telescope + YFOSC + gm3   \\
   11/10/2013  &   6576.25   &     80                    & 3700--8800  &  3          &  2400    & Xinglong 2.16-m telescope + BFOSC + gm4    \\
   29/11/2013  &   6626.03   &    129                    & 3700--8800  &  3          &  3300    & Xinglong 2.16-m telescope + BFOSC + gm4   \\
   06/12/2013  &   6633.05   &    136                    & 3700--8800  &  3          &  3000    & Xinglong 2.16-m telescope + BFOSC + gm4   \\
   27/12/2013  &   6653.99   &    157                    & 3800--8800  &  3          &  3600    & Xinglong 2.16-m telescope + BFOSC + gm4   \\
   \hline
  \end{tabular}
 \end{center}
\tablenotetext{a}{Relative to the estimated date of explosion JD = 2456497.0.}

\end{table}

 \begin{table}
 \caption{Peak Parameters of SN 2013ej in $BVRI$ bands.}  \label{tab:maxpara}
 \begin{center}
  \begin{tabular}{ccccc}
  \hline \hline
                                     &  $B$            &  $V$           &  $R$           &  $I$   \\
   Peak magnitude                    &  12.56$\pm$0.04 & 12.45$\pm$0.05 & 12.24$\pm$0.08 & 12.19$\pm$0.09 \\
   Maximum date\tablenotemark{a}     &  10.23$\pm$0.40 & 15.07$\pm$1.00 & 16.90$\pm$1.00 & 21.49$\pm$1.00  \\
    \hline
  \end{tabular}
 \end{center}
\tablenotetext{a}{Relative to the estimated date of explosion JD = 2456497.0.}
\end{table}

\end{document}